\documentclass[sigplan,twocolumn,nonacm]{acmart}
\renewcommand\footnotetextcopyrightpermission[1]{}
\usepackage{fontspec}
\usepackage{listings}
\usepackage{graphicx}
\usepackage{subcaption}
\usepackage{booktabs}
\usepackage{multirow}
\usepackage{float}
\usepackage{stfloats}

\usepackage{xcolor}
\usepackage{pifont}
\usepackage{wasysym}

\newcommand{\proj}[1]{$Weave$}

\begin{document}
\title[Weave]{Weave: Fine-Grained Dynamic SM Scheduling in an MoE Megakernel for Compute-Communication Overlap}

\author{Ziyu Huang$^{*1}$, Yangjie Zhou$^{*2}$, Chenhao Zhu$^{1}$, Peng Yu$^{3}$, Zihan Liu$^{1}$, Jinyu Liu$^{1}$, Shulai Zhang$^{1}$, Xingxun Tang$^{1}$, Hongzhe Yan$^{1}$, Xinhao Luo$^{1}$, Minyi Guo$^{1}$, Xiu Lin$^{4}$, Yinghao Yu$^{4}$, Guodong Yang$^{4}$, Liping Zhang$^{4}$, Shixuan Sun$^{1}$, Jingwen Leng$^{1}$}
\affiliation{%
  \institution{$^{1}$Shanghai Jiao Tong University}
  \city{Shanghai}
  \country{China}
}
\affiliation{%
  \institution{$^{2}$National University of Singapore}
  \country{Singapore}
}
\affiliation{%
  \institution{$^{3}$Fudan University}
  \city{Shanghai}
  \country{China}
}
\affiliation{%
  \institution{$^{4}$Alibaba Group}
  \country{China}
}
\email{huang_ziyu@sjtu.edu.cn}

\renewcommand{\shortauthors}{Ziyu Huang et al.}

\begin{abstract}
Mixture-of-Experts (MoE) inference under expert parallelism (EP) turns each MoE layer into a distributed computation with costly dispatch and combine communication. State-of-the-art systems reduce this cost through communication-computation overlap, splitting the GPU's SMs for communication and computation respectively. However, this approach still leaves GPU resources wasted along two dimensions. Spatially, the best SM split is determined by each layer's routing result and varies across layers and GPUs, so fixed policies mismatch the workload and waste either NVLink bandwidth or compute throughput. Temporally, complex MoE data dependencies introduce bubbles that leave SMs idle.

We present \textbf{Weave}, to our knowledge the first MoE overlap system that performs fine-grained dynamic SM scheduling---deciding per layer and per GPU by routing results at runtime. Once routing completes, each layer's communication and computation volumes become known; Weave exploits this predictability through a lightweight cost model running inside the persistent megakernel: a spatial scheduler partitions SMs into communication workers and computation workers to match the communication/computation throughput ratio, and a temporal scheduler coordinates the two worker groups to minimize SM idleness. On 4$\times$H100 SXM GPUs across six mainstream MoE models, Weave achieves a $2.89\times$ geometric-mean MoE-layer speedup and a $1.33\times$ geometric-mean end-to-end speedup over five state-of-the-art baselines.
\end{abstract}


\keywords{MoE, Expert Parallelism, communication-\hspace{0pt}computation overlap, GPU SM scheduling}

\maketitle

\begingroup
\renewcommand{\thefootnote}{}
\footnotetext{$^{*}$Both authors contributed equally to this work.}
\endgroup

\section{Introduction}
\label{sec:intro}

Mixture-of-Experts (MoE) scales large language models by replacing dense feed-forward networks with sparsely activated expert networks~\cite{shazeer2017moe}. Each token is routed to only a small subset of experts, increasing model capacity while keeping per-token computation nearly constant. As modern MoE models grow to tens or hundreds of experts~\cite{gshard,switch_transformers,deepseek_moe,olmoe,mixtral,qwen3,qwen35}, distributed inference commonly relies on expert parallelism (EP), where experts are partitioned across GPUs. Under EP, tokens are routed to experts placed on different GPUs, turning each MoE layer into a distributed computation with cross-GPU token movement.

\begin{figure}[t]
\centering
\includegraphics[width=\linewidth]{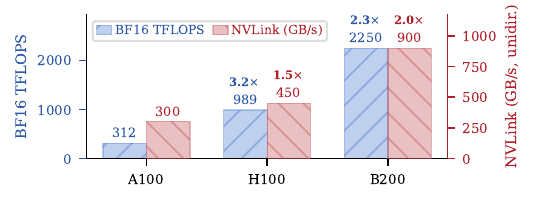}
\caption{GPU compute vs.\ NVLink bandwidth scaling across A100/H100/B200. Left blue axis: BF16 dense TFLOPS. Right red axis: NVLink unidirectional GB/s.}
\label{fig:gpu-scaling}
\end{figure}

This distributed execution makes dispatch and combine a central cost of MoE inference. In each MoE layer, dispatch sends token activations to the GPUs that host their selected experts; after expert computation, combine returns and reduces the expert outputs into final token representations. Prior work reports that dispatch and combine can consume up to 40\% of MoE layer latency~\cite{comet}. This pressure is amplified by hardware scaling: as shown in Figure~\ref{fig:gpu-scaling}, BF16 tensor throughput has grown faster than NVLink bandwidth across recent GPU generations, and this widening compute-interconnect gap makes communication an increasingly important component of MoE layer latency.

\begin{figure}[t]
\centering
\includegraphics[width=\columnwidth,trim=20 1105 24 21,clip]{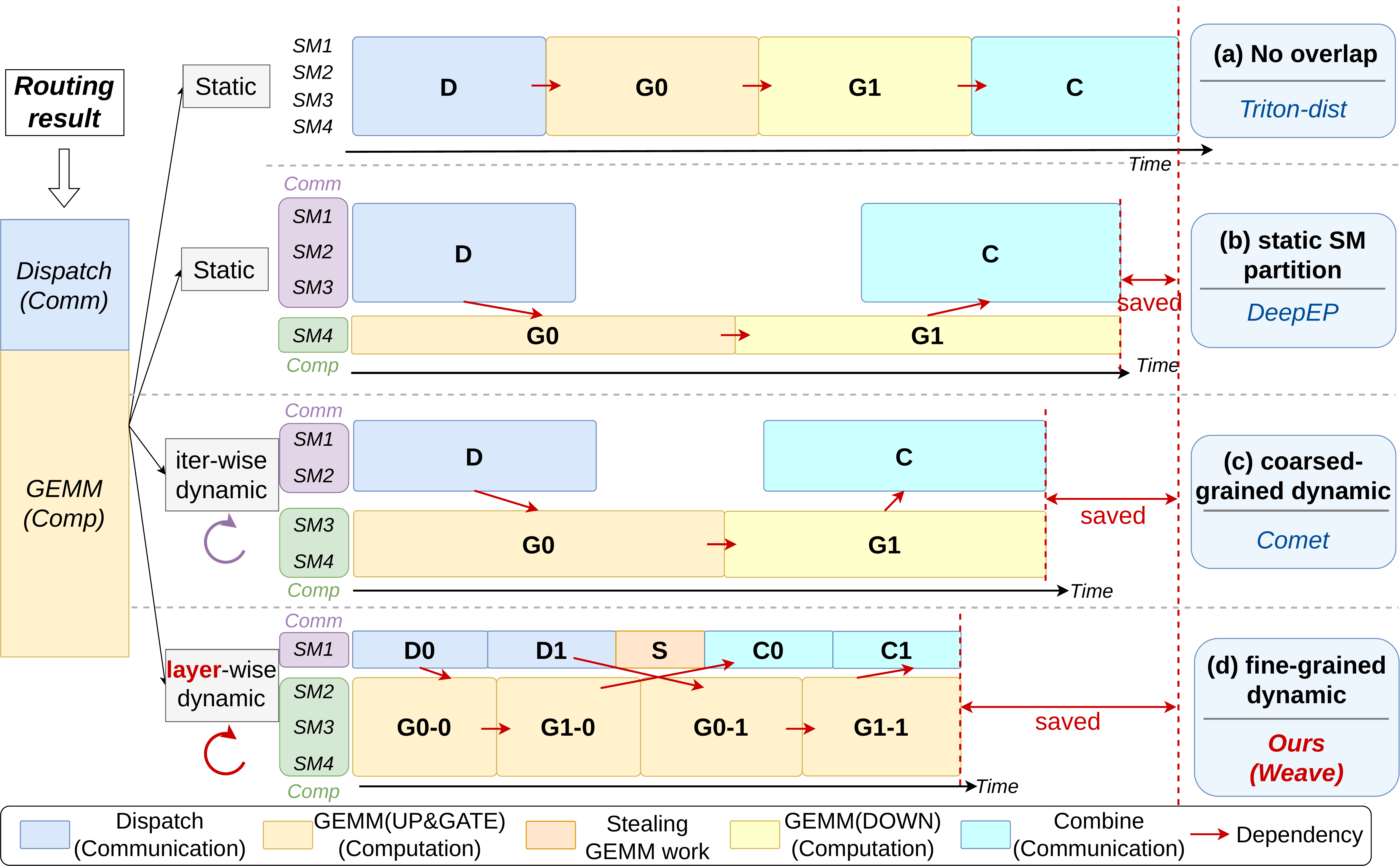}
\caption{Four MoE execution patterns. (a)~No overlap (TD): all SMs execute dispatch, GEMM, and combine serially. (b)~Static SM partition (DeepEP): communication and computation SMs are fixed at compile time, leaving idle bubbles. (c)~Coarse-grained dynamic (Comet): SM count adjusts per iteration. (d)~Fine-grained dynamic (Weave): SM partition adjusts per layer and per GPU.}
\label{fig:intro-overview}
\end{figure}

To reduce this exposed communication cost, software-level communication-computation overlap has become critical for efficient MoE inference. Based on their awareness of the dynamic routing results, existing systems fall into three categories. Figure~\ref{fig:intro-overview} compares the execution timelines of these approaches.

\textbf{(i)~No explicit SM partitioning} Triton-Distributed (TD)~\cite{tritondist} runs communication and computation serially across all SMs, relying on global preemption for overlap at operator boundaries. All SMs execute dispatch, GEMM, and combine in sequence, yielding minimal communication-computation overlap.

\textbf{(ii)~Static SM partition} DeepEP~\cite{deepep} and ParallelKittens (PK)~\cite{parallelkittens} fix the number of communication SMs at compile time, applying the same partition to all layers. Dedicating separate SM groups to communication and computation enables overlap, but data dependencies in the communication-computation pipeline (up gemm cannot begin until dispatch delivers tokens; combine must wait for down gemm to finish) create pipeline bubbles that leave SMs idle.

\textbf{(iii)~Coarse-grained dynamic} Comet~\cite{comet} selects the communication/computation SM ratio from a pre-profiled kernel library based on the input sequence length, switching once per iteration. This adapts to workload intensity across iterations, but the partition still does not respond to per-layer routing variation.

However, none of these approaches adapts the SM partition to the actual routing result of each layer, leaving substantial GPU resource waste along both spatial and temporal dimensions.

The waste manifests in two dimensions. Spatially, the optimal communication/computation SM ratio depends on each layer's routing result, which determines the communication and computation volumes on each GPU. When the SM partition does not match these volumes, one group of SMs finishes early and idles while the other remains the bottleneck, wasting either communication bandwidth or computation throughput. Since routing skew varies across inputs, layers, and GPUs, a fixed or per-iteration partition inevitably mismatches some layers.

Temporally, even with a well-chosen SM split, the pipeline bubbles shown in Figure~\ref{fig:intro-overview}(b) persist: communication SMs remain idle during the middle computation phase, forming a large \emph{mid bubble} that further degrades SM utilization.

Building on these observations, we propose \textbf{Weave}, a runtime SM scheduling system for MoE communication-computation overlap (Figure~\ref{fig:intro-overview}(d)). Weave exploits the fact that each layer's communication and computation volumes become known after routing, and schedules SMs per layer and per GPU rather than relying on an offline fixed policy. It combines a spatial scheduler that partitions SMs into communication workers and computation workers with a temporal scheduler that coordinates the two worker groups to minimize SM idleness, unified by a lightweight cost model running inside a persistent megakernel.

We implement Weave on PK~\cite{parallelkittens} and evaluate it on 4$\times$H100 SXM GPUs across six mainstream MoE architectures. Compared with five state-of-the-art baselines~\cite{sglang,parallelkittens,tritondist,deepep,comet}, Weave achieves the lowest per-layer MoE latency in all configurations, with a $2.89\times$ geometric-mean MoE-layer speedup and a $1.33\times$ geometric-mean end-to-end speedup.

This paper makes the following contributions.
\begin{itemize}
  \item We identify routing-determined SM scheduling as a key inefficiency in distributed MoE inference, showing that the communication/computation SM split must be selected per layer and per GPU after routing completes.

  \item We design Weave's runtime spatial scheduler, which uses routing-derived workload volumes and calibrated hardware throughput curves to choose the communication SM count for each layer and GPU.

  \item We design Weave's temporal scheduler, which uses chunk pipelining and bubble stealing to reduce SM idleness caused by MoE data dependencies within each layer.

  \item We implement Weave based on Parallel Kittens~\cite{parallelkittens} and show that it consistently improves MoE-layer and end-to-end inference latency over state-of-the-art baselines.
\end{itemize}

\noindent Our implementation will be open-sourced upon acceptance.

\section{Background and Motivation}
\label{sec:motivation}
This section first introduces the basics of Mixture-of-Experts (MoE) and the corresponding computation/communication overlap techniques.
Then we analyze the limitations of existing communication-related optimizations to derive our insights.

\subsection{Mixture-of-Experts and Expert Parallelism}
\label{sec:bg-moe}
MoE replaces the monolithic FFN layer with tens to hundreds of smaller expert sub-networks, each activated only for a subset of tokens selected by a learned gating mechanism, thereby scaling model parameters while keeping per-token computation nearly constant.

The presence of multiple discrete experts gives rise to a new parallelism paradigm: expert parallelism (EP), under which experts are partitioned across GPUs. Hosting experts across multiple GPUs introduces cross-GPU communication overhead.

\begin{figure}[htbp]
  \centering
  \includegraphics[width=\columnwidth]{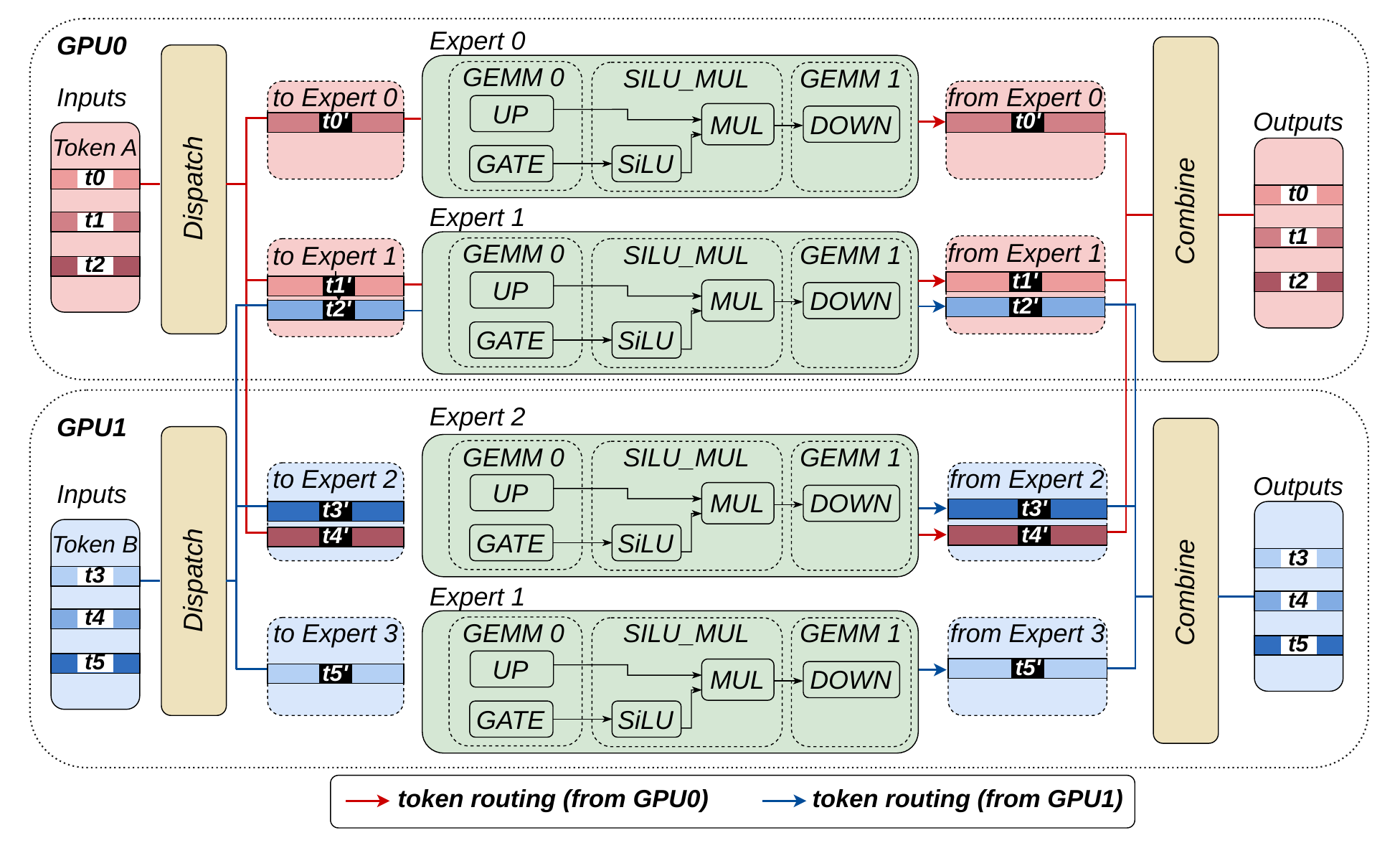}
  \caption{MoE forward pass under expert parallelism (2~GPUs, 4~experts, top-$k$\,=\,3 routing). Dispatch sends each token to the GPU hosting its expert; the expert sub-network sequentially executes gemm0 (UP+GATE), silu\_mul, and gemm1 (DOWN); combine performs weighted summation and returns the results.}
  \label{fig:moe-forward}
\end{figure}

As illustrated in Figure~\ref{fig:moe-forward}, each MoE layer under EP consists of five stages: (1)~\textbf{\emph{dispatch}}, which sends each token to the GPU hosting its target experts; (2)~\textbf{\emph{GEMM0}}, the up-projection and gated FFN; (3)~\textbf{\emph{activation}}, typically a SiLU followed by element-wise multiplication; (4)~\textbf{\emph{GEMM1}}, the down-projection FFN; and (5)~\textbf{\emph{combine}}, which aggregates the per-expert results via weighted summation and returns them to the originating GPU. Among these, dispatch and combine are cross-GPU communication operations, while GEMM0, activation, and GEMM1 are local expert computation.

\subsection{Compute-Communication Overlap}
\label{sec:bg-overlap}
To hide dispatch and combine latency, modern systems overlap communication with computation by partitioning GPU SMs between the two tasks; for example, DeepEP~\cite{deepep} dedicates 20 of the 132 SMs on an H100 to communication and uses the remaining 112 for computation.

\begin{figure*}[htbp]
\centering
\includegraphics[width=0.88\textwidth]{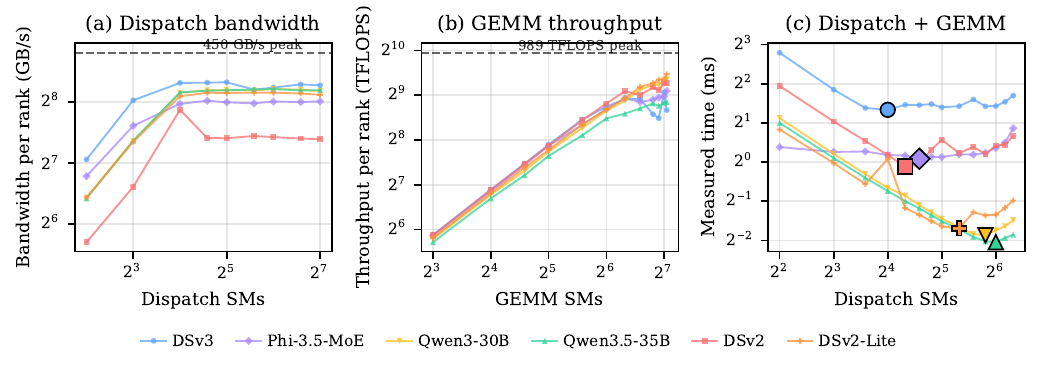}
\caption{Impact of SM partitioning on compute-communication overlap (4$\times$H100 SXM, 6 MoE models, seq\,=\,8192 prefill, EP\,=\,4, BF16). (a)~Dispatch bandwidth vs.\ dispatch SM count $c$: near-linear then saturates. (b)~GEMM throughput vs.\ compute SM count $N{-}c$: sub-linear under power and L2 cache constraints. (c)~Dispatch-gemm total time with tile-level overlap as a function of $c$; markers indicate each model's optimal split.}
\label{fig:sm-constraint}
\end{figure*}

However, even SM-level partitioning is a non-trivial design problem. We conduct an experiment that evaluates different compute-communication SM partitions across six MoE models (DeepSeek-V3 (DSv3), Phi-3.5-MoE, Qwen3-30B, Qwen3.5-35B, DeepSeek-V2-Lite (DSv2-Lite), DeepSeek-V2 (DSv2); seq\,=\,8192 prefill, 4$\times$H100 SXM, EP\,=\,4, BF16). As shown in Figure~\ref{fig:sm-constraint}(c), on the one hand, the optimal partition point that minimizes overall execution time varies across models, indicating that the SM partition must be adapted per layer. On the other hand, the trend itself also differs across models, suggesting that the strategy should adapt to different models, configurations, and other factors.

\subsection{Limitations of SOTA Approaches}
\label{sec:bg-limitations}

Existing distributed MoE systems either fix the number of communication SMs at compile time~\cite{deepep,parallelkittens}, select from a pre-profiled configuration per iteration~\cite{comet}, or avoid explicit SM partitioning altogether and run communication and computation serially~\cite{tritondist}. However, MoE workloads are \emph{inherently dynamic}: the routing result of each layer changes, causing the communication and computation volumes to shift, and consequently the optimal communication SM count varies from layer to layer. Resolving this mismatch poses two key challenges: how to determine the optimal SM partition at runtime, and, once the partition is made, how to coordinate the two groups of SMs (comm workers and comp workers) to maximize overlap while minimizing idle time.

\subsubsection{Challenge A: Runtime Partition Decision}
\label{subsubsec:challenge-a}
\leavevmode\\
\begin{figure}[!t]
  \centering
  \includegraphics[width=\linewidth]{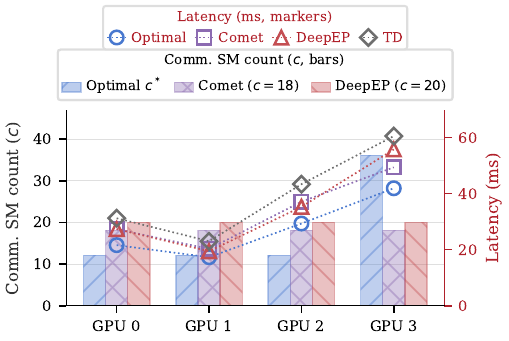}
  \caption{SM partition mismatch under existing policies (DSv2, CodeSearchNet dataset, Layer~1, 4$\times$H100 SXM). Left y-axis (hatched bars): per-GPU optimal comm SM count $c^*$ vs.\ Comet ($c{=}18$, per-iteration) and DeepEP ($c{=}20$, static); the optimal $c^*$ ranges from 12 to 36 across GPUs. Right y-axis (hollow markers): per-GPU latency under each policy plus TD (serial); layer latency is limited by the slowest GPU.}
  \label{fig:spatial-motivation}
\end{figure}

As discussed in \S\ref{sec:intro}, existing approaches fall into three suboptimal modes: static fixed, coarse-grained dynamic, and no overlap. We now quantify the performance gap on DSv2 (4$\times$H100 SXM) with CodeSearchNet dataset. As the hatched bars of Figure~\ref{fig:spatial-motivation} show, within a single MoE layer the per-GPU optimal $c^*$ ranges from 12 to 36. Neither DeepEP's fixed $c{=}20$ nor Comet's per-iteration $c{=}18$ matches any GPU's optimum. Since all GPUs synchronize after each MoE layer, the layer latency is determined by the slowest GPU. The hollow markers report the resulting per-GPU latency: on the bottleneck GPU (GPU~3), DeepEP is $1.33\times$ slower than the per-GPU optimal, Comet $1.18\times$, and TD $1.44\times$.

Determining the optimal SM partition at runtime is therefore a key challenge.

\subsubsection{Challenge B: Dynamic Coordination of Two Worker Groups}
\label{subsubsec:challenge-b}
\leavevmode\\
Even after performing the SM partition described in Challenge~A, complex dependencies between communication and computation tasks make it difficult to coordinate the two worker groups so that they run in parallel (high comm/comp overlaping) without waiting for each other (high SM Active rate).

\begin{figure}[!t]
  \centering
  \includegraphics[width=\linewidth]{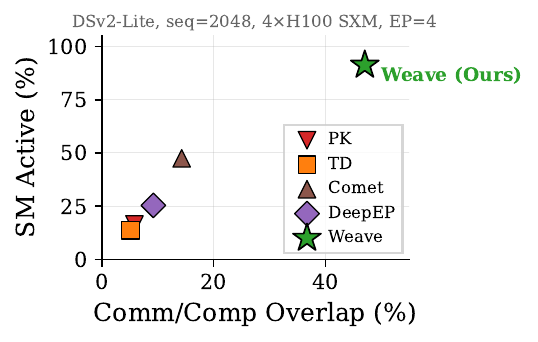}
  \caption{Performance of MoE overlap systems along two utilization dimensions: comm-comp overlap (x-axis) and SM Active (y-axis). Each marker is one system, measured on DSv2-Lite (seqlen\,=\,2048, 4$\times$H100 SXM, EP\,=\,4, BF16).}
  \label{fig:2d-scatter}
\end{figure}

The MoE forward pass has inherent data dependencies: up gemm cannot begin until dispatch delivers the required tokens, and combine cannot start until down gemm finishes. Under SM partitioning, these dependencies cause communication workers to idle during the middle computation phase, forming pipeline bubbles that existing systems~\cite{deepep,parallelkittens,comet} do not fill. As shown in Figure~\ref{fig:2d-scatter}, no existing system simultaneously achieves high SM Active and high comm-comp overlap; only Weave maintains high levels on both dimensions (91\% SM Active, 47\% overlap).

\section{Design}
\label{sec:design}

\begin{figure}[!t]
  \centering
  \includegraphics[width=\linewidth,trim=0 140 0 0,clip]{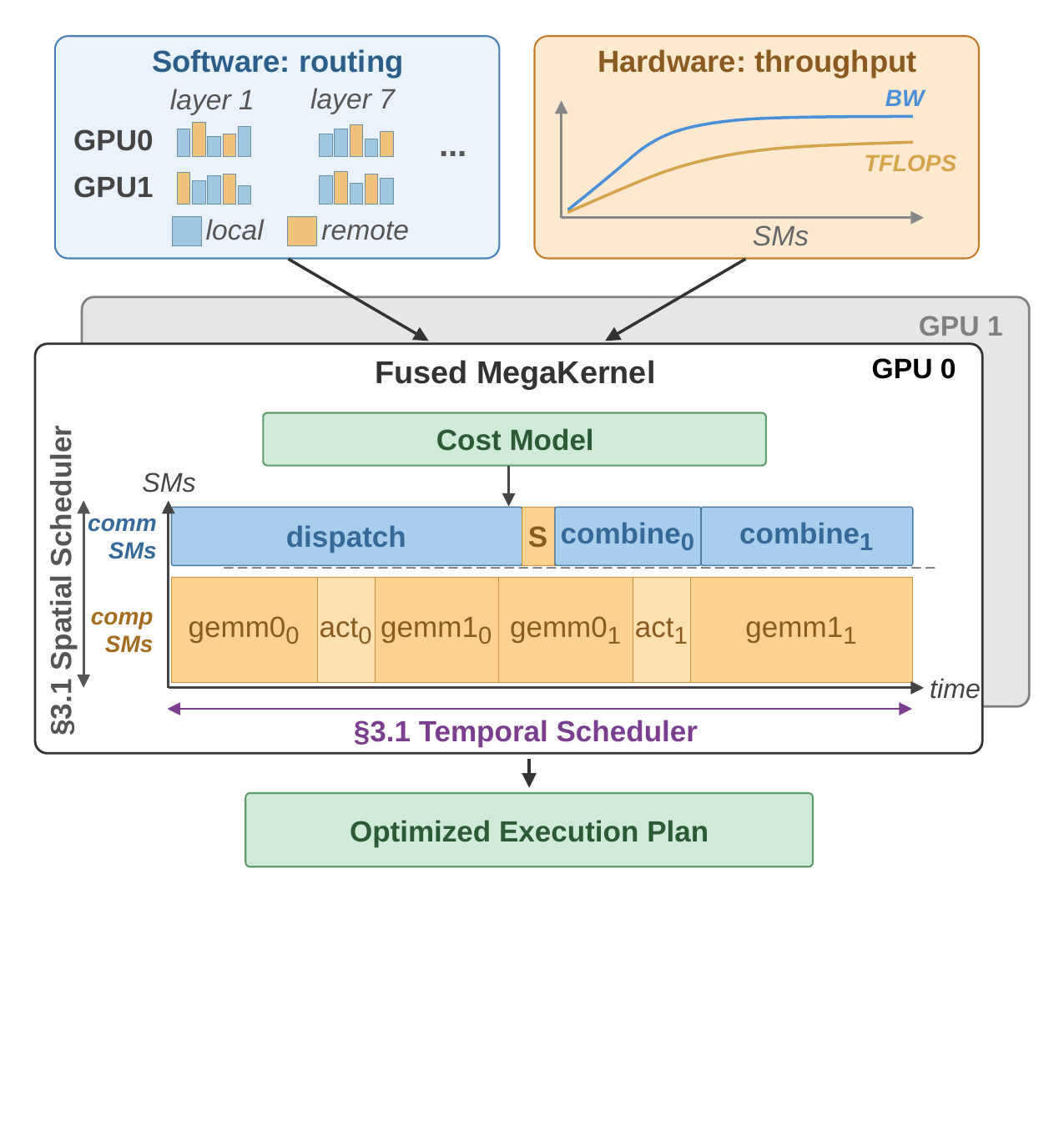}
  \caption{System overview of Weave. The system takes routing results (software input) and communication bandwidth / compute throughput curves (hardware input), and makes runtime decisions inside the megakernel along two dimensions: spatial scheduler (how to split SMs between communication and computation, \S\ref{subsec:spatial}), and temporal scheduler (how the two types of workers coordinate their execution, \S\ref{subsec:temporal}). The output is an optimized execution plan applied to the current layer.}
  \label{fig:weave-overview}
\end{figure}

In this section, we present the core design of Weave, a runtime SM scheduling system for efficient MoE execution under expert parallelism. Existing systems cannot perceive per-layer routing results and adapt the SM partition accordingly, which leads to missed optimization opportunities in two dimensions: spatially, a mismatched SM split wastes either communication bandwidth or compute throughput; temporally, the complex dependencies introduced by communication-computation overlap leave SMs idle. Weave addresses both dimensions with runtime scheduling.

Figure~\ref{fig:weave-overview} gives an overview of Weave's architecture. Weave takes two categories of input: software input, consisting of per-layer routing results that determine the communication volume and computation workload on each GPU; and hardware input, consisting of calibrated communication bandwidth and compute throughput curves as functions of SM count. These inputs are fed into a persistent megakernel that fuses all five forward operations of an MoE layer (dispatch, gemm0, silu\_mul, gemm1, combine). Within this megakernel, each SM is assigned either a \emph{comm} role (Dispatch and Combine) or a \emph{comp} role (GEMM0, SiLU-Mul, and GEMM1). A lightweight cost model in the megakernel prologue jointly determines the spatial SM partition and the temporal execution order, producing a per-layer, per-GPU scheduling plan that coordinates the two types of workers throughout the layer.

The key observation behind Weave's runtime scheduling is that the workload of an MoE layer becomes fully determined once routing completes. At that point, Weave knows the number of tokens to send and receive, the token distribution across experts, and the resulting computation volume on each GPU. Weave therefore invokes its scheduler immediately after routing. The scheduler combines these runtime workload descriptors with hardware throughput profiles and uses a lightweight cost model inside the megakernel prologue to generate a layer-specific scheduling plan within microseconds. The megakernel then transitions to its main execution phase.

Weave jointly optimizes scheduling along two dimensions. The spatial scheduler (\S\ref{subsec:spatial}) selects the number of SMs assigned to communication so that communication bandwidth and computation throughput are balanced for the current layer. The temporal scheduler (\S\ref{subsec:temporal}) chunks routed tokens into pipelined microbatches and allows idle communication SMs to steal GEMM tiles when communication work is temporarily unavailable. Since the best temporal schedule depends on the throughput induced by the spatial SM split, Weave solves the two dimensions jointly using a unified cost model (\S\ref{subsec:joint-search}).

Concretely, the scheduling decision is governed by two variables: the number of communication SMs $c$ (how to partition SMs between the two worker groups) and the pipeline chunk count $K$ (how to split tokens into pipelined microbatches for temporal coordination). The cost model (\S\ref{subsec:joint-search}) jointly searches over the $(c, K)$ space and, for each candidate pair, derives the number of GEMM tiles $n_\text{steal}$ that communication workers can steal during their idle periods. The optimal $(c^*, K^*)$ and the corresponding steal count fully determine the execution plan for each layer and GPU.

\subsection{Spatial Scheduler}
\label{subsec:spatial}

The spatial scheduler determines how the available SMs are divided between communication and computation. As shown in Figure~\ref{fig:sm-constraint}(c), the overall latency as a function of the SM split has a clear optimum that depends on the routed workload of the current layer.

As shown in Figure~\ref{fig:sm-constraint}(a)(b), both communication bandwidth and GEMM throughput saturate well before all $N$ SMs are assigned; a small number of communication SMs is sufficient to approach peak bandwidth, leaving the majority for computation.


\begin{figure}[!t]
  \centering
  \includegraphics[width=0.98\linewidth]{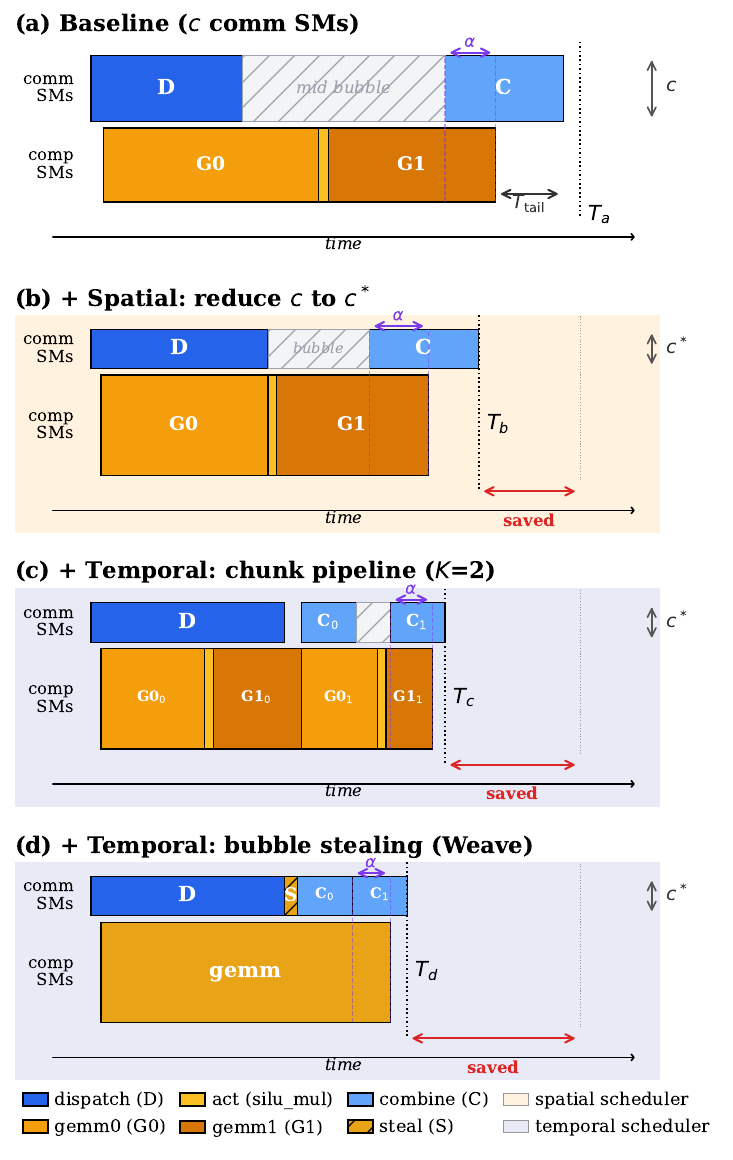}
  \caption{Operational details of the spatial and temporal schedulers. (a)~Baseline with communication-computation overlap enabled; a large mid bubble persists. (b)~The spatial scheduler applies a near-optimal communication/computation SM ratio, reducing the mid bubble. (c)~The temporal scheduler partitions tokens into microbatches, exposing additional overlap opportunities and further reducing the mid bubble. (d)~The temporal scheduler fills the residual mid bubble by assigning idle communication SMs to execute GEMM tiles.}

  \label{fig:progressive}
\end{figure}

We now formalize the optimization. Let $N$ denote the number of SMs available to the megakernel, and let $c$ denote the number of SMs assigned to communication; the remaining $N{-}c$ SMs are assigned to computation. Let $H$ denote the hidden dimension, $I$ the per-expert FFN intermediate dimension, and $B = 2H$\,bytes the per-token communication size (each token is an $H$-element BF16 vector, 2\,bytes per element). After routing, let $X_{\text{local}}$ denote the number of local tokens routed to local experts, $X_{\text{in}}$ the number of remote tokens routed to local experts, and $X_{\text{in}}^{\text{uniq}}$ the deduplicated count of $X_{\text{in}}$ (under top-$k$ routing with $k{>}1$, the same remote token may be selected by multiple local experts, so $X_{\text{in}}^{\text{uniq}} \le X_{\text{in}}$; each unique remote token is transferred across the network only once and then reused for all selecting experts~\cite{deepep, uniep}). The per-layer workloads are:
\begin{align}
W_{\text{comp}} &= (X_{\text{local}} + X_{\text{in}}) \cdot 6HI \;\text{FLOPs}, \label{eq:wcomp} \\
W_{\text{dispatch}} &= X_{\text{in}}^{\text{uniq}} \cdot B \;\text{bytes}, \label{eq:wdispatch} \\
W_{\text{combine}} &= X_{\text{in}} \cdot B \;\text{bytes}. \label{eq:wcombine}
\end{align}
$W_{\text{comp}}$ counts the GEMM work executed jointly on $X_{\text{local}} + X_{\text{in}}$ token-expert pairs (both local and remote). For communication, the dispatch volume is $X_{\text{in}}^{\text{uniq}} \cdot B$ (each unique remote token is pulled once and reused) and the combine volume is $X_{\text{in}} \cdot B$ (one per-expert output returned for each token-expert pair).

We let $\alpha \in [0, 1)$ denote the tile-level overlap ratio between communication and computation, calibrated empirically. The remaining fraction $(1 - \alpha)$ of communication is exposed as a tail outside the overlap window. The three time components are
\begin{align}
T_{\text{comp}} &= \frac{W_{\text{comp}}}{\text{TFLOPS}(N{-}c)}, \label{eq:tcomp} \\
T_{\text{comm}} &= \frac{W_{\text{dispatch}} + W_{\text{combine}}}{\text{BW}(c)}, \label{eq:tcomm} \\
T_{\text{tail}} &= (1 - \alpha) \cdot \frac{W_{\text{combine}}}{\text{BW}(c)}. \label{eq:ttail}
\end{align}
The layer is bounded by whichever of the two paths is slower, plus the unhidden tail. The spatial scheduler therefore selects
\begin{align}
T_{\text{total}} &= \max\!\big(T_{\text{comp}} + T_{\text{tail}},\; T_{\text{comm}}\big), \label{eq:ttotal} \\
c^* &= \arg\min\; T_{\text{total}}. \label{eq:cstar}
\end{align}
If $T_{\text{comp}} + T_{\text{tail}} > T_{\text{comm}}$, the layer time is determined by the GEMM together with the unhidden combine tail; otherwise the GEMM fits entirely within the communication window and the layer time is determined by serialized dispatch and combine.

\subsection{Temporal Scheduler}
\label{subsec:temporal}

Given the SM partition $c$ from the spatial scheduler, the temporal scheduler addresses how the two groups of SMs coordinate over time. As Figure~\ref{fig:progressive}(a) shows, under a fixed SM partition with no temporal scheduling, Dispatch and Combine cluster at the two ends of the timeline, leaving a large stretch of idle comm SMs in the middle of the MoE layer and forming a large \emph{mid bubble}, i.e., an idle interval during the GEMM phase.

To eliminate this bubble, Weave uses two mechanisms: the \textbf{chunk pipeline} (\S\ref{subsubsec:chunk-pipeline}) spreads Combine into the middle of the GEMM phase for fine-grained overlap; \textbf{bubble stealing} (\S\ref{subsubsec:bubble-steal}) lets comm SMs steal GEMM tiles in the residual mid bubble that the chunk pipeline still cannot cover.

\subsubsection{Chunk Pipeline}
\label{subsubsec:chunk-pipeline}
\leavevmode\\
The mid bubble in Figure~\ref{fig:progressive}(a) comes from the DAG dependency: a Combine operation cannot start until the corresponding GEMM1 tile has been produced. To relax this constraint, we split the tokens into smaller chunks; chunks are independent of each other, which opens up more overlap opportunities. As Figure~\ref{fig:progressive}(c) shows, we split GEMM0, SiLU-Mul, GEMM1, and Combine into $K$ chunks, while Dispatch is not chunked, so that communication can overlap with computation as much as possible. Unlike a naive approach where comm SMs switch to comp after dispatch finishes, the chunk pipeline keeps comm SMs continuously doing communication throughout the middle phase. Figure~\ref{fig:progressive}(c) illustrates $K{=}2$: chunk 0's Combine ($C_0$) runs during chunk 1's GEMM, so communication and computation execute concurrently in time.

The choice of $K$ involves a trade-off: a larger $K$ gives finer chunks and a smaller mid bubble, but each chunk holds fewer tokens, degrading per-chunk GEMM throughput. As Figure~\ref{fig:chunk-k} shows, we benchmark the group GEMM performance under the chunk pipeline on H100 using DSv3 with input seq\,=\,8k. As $K$ increases, the performance degradation accelerates; at $K{=}8$, throughput drops by ${\sim}27\%$ compared to $K{=}1$. We capture this degradation with a correction factor $\text{eff}_{\text{chunk}}(K) \in (0, 1]$, defined as the ratio of chunked GEMM throughput to unchunked throughput ($K{=}1$). The computation time under chunking becomes:
\begin{equation}
T_{\text{comp}}(c, K) = \frac{W_{\text{comp}}}{\text{TFLOPS}(N{-}c) \cdot \text{eff}_{\text{chunk}}(K)}, \label{eq:tcomp-chunked}
\end{equation}
where $\text{eff}_{\text{chunk}}(K) \in (0, 1]$ is the ratio of chunked GEMM throughput to the unchunked case. Chunking also modifies the tail term: the final chunk's combine, the only segment not absorbed by a subsequent GEMM, is now of size $W_{\text{combine}}/K$, so the tail is reduced by a factor of $K$ relative to the unchunked case:
\begin{equation}
T_{\text{tail}} = (1 - \alpha) \cdot \frac{W_{\text{combine}}}{\text{BW}(c) \cdot K}. \label{eq:ttail-chunked}
\end{equation}
A larger $K$ reduces the unhidden tail but increases the GEMM time through the degradation of $\text{eff}_{\text{chunk}}(K)$; a smaller $K$ does the opposite. The optimum balances these two effects.


\subsubsection{Bubble Stealing}
\label{subsubsec:bubble-steal}
\leavevmode\\
After the joint search of \S\ref{subsec:joint-search} chooses $(c^*, K^*)$, communication SMs may still have idle capacity between the end of dispatch and the start of the chunked combine. We fill this gap by letting them steal GEMM tiles, shown as the S block in Figure~\ref{fig:progressive}(d). Once Dispatch finishes, comm SMs join the $N{-}c$ comp SMs on GEMM work, converting idle capacity into effective GEMM throughput. We place the steal window right after Dispatch so that stolen GEMM tiles execute consecutively and subsequent Combine operations also run without interruption. If stealing were instead scattered into the idle gaps between each chunk's Combine, both GEMM and Combine would be repeatedly interrupted by role switches, degrading performance. This consolidated placement yields the pattern shown in Figure~\ref{fig:progressive}(d).

Below we derive the number of tiles to steal. Given $c$, communication takes
\begin{equation}
t_{\text{comm}}(c) = \frac{W_{\text{comm}}}{\text{BW}(c)}. \label{eq:tcomm-stealref}
\end{equation}
During this interval, the $N{-}c$ comp SMs complete
\begin{equation}
W_{\text{partial}}(c) = t_{\text{comm}}(c) \cdot \text{TFLOPS}(N{-}c)
\end{equation}
of GEMM work, leaving
\begin{equation}
W_{\text{steal}}(c) = \max\!\big(0,\; W_{\text{comp}} - W_{\text{partial}}(c)\big) \label{eq:wsteal}
\end{equation}
GEMM FLOPs to be stolen. Once Dispatch finishes, all $N$ SMs are available for GEMM, so $W_{\text{steal}}(c)$ is shared equally across $N$ SMs. Letting $W_{\text{tile}}$ denote the FLOPs of a single GEMM tile, each SM steals
\begin{equation}
n_{\text{steal}}(c) = \frac{W_{\text{steal}}(c)}{N \cdot W_{\text{tile}}} \label{eq:nsteal}
\end{equation}
GEMM tiles.


\subsection{Joint Search for $(c^*, K^*)$}
\label{subsec:joint-search}

\begin{figure}[!t]
\centering
\begin{subfigure}[t]{0.48\linewidth}
  \centering
  \includegraphics[width=\linewidth]{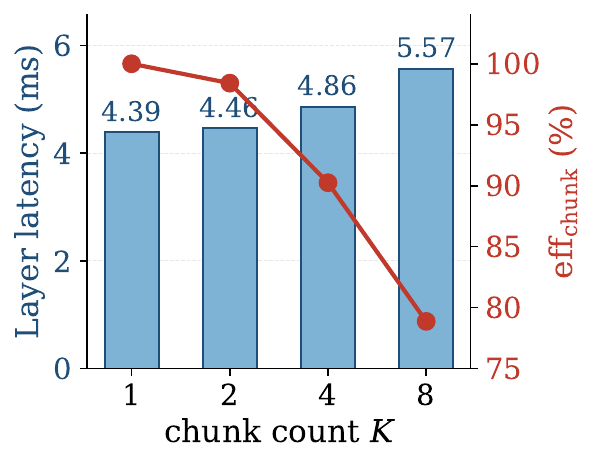}
  \caption{Impact of chunk splitting on GEMM (DSv3, seq=8192, H100).}
  \label{fig:chunk-k}
\end{subfigure}\hfill
\begin{subfigure}[t]{0.48\linewidth}
  \centering
  \includegraphics[width=\linewidth]{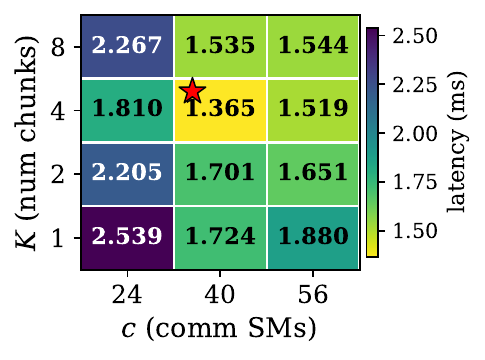}
  \caption{$(c, K)$ latency grid for DSv2-Lite (seq=8192). Star: optimum $c^*\!=\!40, K^*\!=\!4$.}
  \label{fig:ck-contour}
\end{subfigure}
\caption{(a) Chunk splitting degrades GEMM throughput; at $K{=}8$, performance drops by ${\sim}27\%$. (b) Measured latency over the $(c, K)$ grid shows a clear unimodal optimum, confirming joint search is necessary.}
\label{fig:chunk-k-and-contour}
\end{figure}

Substituting the $K$-aware definitions of $T_{\text{comp}}$ and $T_{\text{tail}}$ from \S\ref{subsec:temporal} into the total layer time expression of \S\ref{subsec:spatial} yields
\begin{equation}
T_{\text{total}}(c, K) = \max\!\big(T_{\text{comp}}(c, K) + T_{\text{tail}}(c, K),\; T_{\text{comm}}(c)\big). \label{eq:ttotal-joint}
\end{equation}
The scheduler then jointly searches for
\begin{equation}
(c^*, K^*) = \arg\min_{c, K}\; T_{\text{total}}(c, K). \label{eq:ckstar}
\end{equation}
In the comp-dominated regime(if $T_{\text{comp}} + T_{\text{tail}} > T_{\text{comm}}$), increasing $K$ raises $T_{\text{comp}}$ through the degradation of $\text{eff}_{\text{chunk}}(K)$ but lowers $T_{\text{tail}}$ via the $1/K$ scaling. The two opposing effects produce the real measured $(c, K)$ grid result(Fig.~\ref{fig:ck-contour}). The number of GEMM tiles to steal is then computed from $(c^*, K^*)$.

Figure~\ref{fig:ck-contour} shows the measured latency under different $(c, K)$ configurations. We use DSv2-Lite with input sequence length 8192 for this experiment. The grid exhibits a single clear optimum at $c^*{=}40, K^*{=}4$ (1.365\,ms), confirming that $c$ and $K$ must be searched jointly.

Once $(c^*, K^*)$ is determined, the scheduler further computes the per-SM steal-tile count $n_{\text{steal}}(c^*)$ via equations~\eqref{eq:wsteal} and~\eqref{eq:nsteal} from \S\ref{subsubsec:bubble-steal}. At this point the megakernel's complete execution plan (SM role assignment, chunk count, and the per-SM steal-tile count) is fully determined, and the kernel enters its main execution phase.

\section{Implementation}
\label{sec:impl}

Weave is built on PK~\cite{parallelkittens} and implemented as a single persistent megakernel, launched with a grid size equal to the SM count ($N{=}132$ on H100). After each layer's routing completes, the routing result is passed as a kernel parameter; the megakernel then runs the cost model of \S\ref{subsec:joint-search} on-GPU to obtain the current layer's $(c^*, K^*)$. Both $c^*$ and $K^*$ can therefore switch dynamically per layer.

Each block derives its role from $c^*$ and \texttt{blockIdx.x}: blocks with index ${<}\,c^*$ are comm SMs (executing dispatch and combine), while the remaining blocks are omp SMs (executing gemm0, activation, and gemm1). All tile assignments use dynamic scheduling: each SM claims its next tile via \texttt{atomicAdd} on a global counter following the chunk pipeline DAG order, naturally tolerating variation in SM execution speeds~\cite{thomas2016core}. After dispatch finishes, comm SMs steal $n_{\text{steal}}$ GEMM tiles before transitioning to combine. For the final combine, all $N$ SMs participate regardless of role, eliminating tail idleness.

\label{subsec:dedup}

Weave also fuses dispatch-side token deduplication~\cite{deepep} into the megakernel. DeepEP~\cite{deepep} and TD~\cite{tritondist} perform deduplication as a separate operation before GEMM0, whereas Weave overlaps dispatch with GEMM0 and achieves deduplication naturally during dispatch. After routing completes, duplicate tokens are identified and each comm tile knows in advance whether its token is a duplicate. If so, the tile reads directly from a local HBM buffer populated by the first transfer, avoiding both redundant cross-GPU transfers and HBM polling.

\section{Evaluation}
\label{sec:eval}

\subsection{Evaluation Setup}
\label{subsec:evaluation-setup}

\noindent\textbf{Hardware.}
All experiments are conducted on a node with $4\times$ NVIDIA H100 80\,GB SXM5 GPUs. Each GPU has 132 SMs, 80\,GB HBM3, and a peak BF16 throughput of 989 TFLOPS. The GPUs are fully connected through NVLink and NVSwitch, with 450\,GB/s unidirectional bandwidth per GPU. The host has two Intel Xeon Platinum 8468 CPUs, each with 48 physical cores and 96 hardware threads, for a total of 96 cores and 192 hardware threads across 2 NUMA nodes. The system has 2.0\,TiB of CPU memory.

\noindent\textbf{Software.}
We use CUDA Toolkit 12.9.86, NVIDIA driver 575.57.08, NCCL 2.21.5, NVSHMEM 3.6.5, Python 3.10.12, and PyTorch 2.6.0+cu124. The evaluated baseline implementations are SGLang v0.5.9, Triton-Distributed(TD) v3.4.0, DeepEP v1.2.1, Comet with Flux v1.1.2, and ParallelKittens at commit \texttt{a8f63a9}.

\noindent\textbf{Models.}
We evaluate six mainstream BF16 MoE architectures under expert parallelism with $\mathrm{EP}{=}4$, as summarized in Table~\ref{tab:models}. Here, $E$ denotes the total number of routed experts, top-$k$ denotes the number of experts selected per token, $H$ denotes the hidden dimension, and $I$ denotes the per-expert intermediate dimension.

\begin{table}[!t]
\centering
\caption{MoE model configurations used in evaluation.}
\label{tab:models}
\small
\begin{tabular}{lcccc}
\toprule
\textbf{Model} & $E$ & \textbf{top-}$k$ & $H$ & $I$ \\
\midrule
DSv3           & 256 & 8 & 7168 & 2048 \\
Phi-3.5-MoE    &  16 & 2 & 4096 & 6400 \\
Qwen3-30B      & 128 & 8 & 2048 &  768 \\
Qwen3.5-35B    & 256 & 8 & 2048 &  512 \\
DSv2-Lite      &  64 & 6 & 2048 & 1408 \\
DSv2           & 160 & 6 & 5120 & 1536 \\
\bottomrule
\end{tabular}
\end{table}

\noindent\textbf{Workloads.}
We sample routing inputs from the ShareGPT dataset and control the input sequence length in $\{2048,4096,8192\}$ with batch size 1.

\noindent\textbf{Baselines.}
We compare Weave with five representative MoE serving and communication-computation overlap systems. SGLang~\cite{sglang} is a production-grade serving system that uses NCCL All-to-All and cuBLAS grouped GEMM. TD~\cite{tritondist} fuses dispatch with GEMM0 and GEMM1 with combine, and uses global task preemption within fused operators to reduce inter-operator bubbles. DeepEP~\cite{deepep} is an expert-parallel communication library that chunks tokens into pipeline stages to overlap dispatch and combine with expert GEMM; we use DeepGEMM~\cite{deepgemm} as its GEMM backend and keep its default communication partition of 20 SMs. Comet~\cite{comet} fuses dispatch with GEMM0 and GEMM1 with combine, enabling tile-level communication-computation overlap. ParallelKittens (PK)~\cite{parallelkittens} fuses dispatch with expert up-projection under a compile-time fixed SM partition.

\subsection{Per-Layer MoE Latency}
\label{subsec:main-results}

\begin{figure*}[htbp]
\centering
\includegraphics[width=0.95\textwidth]{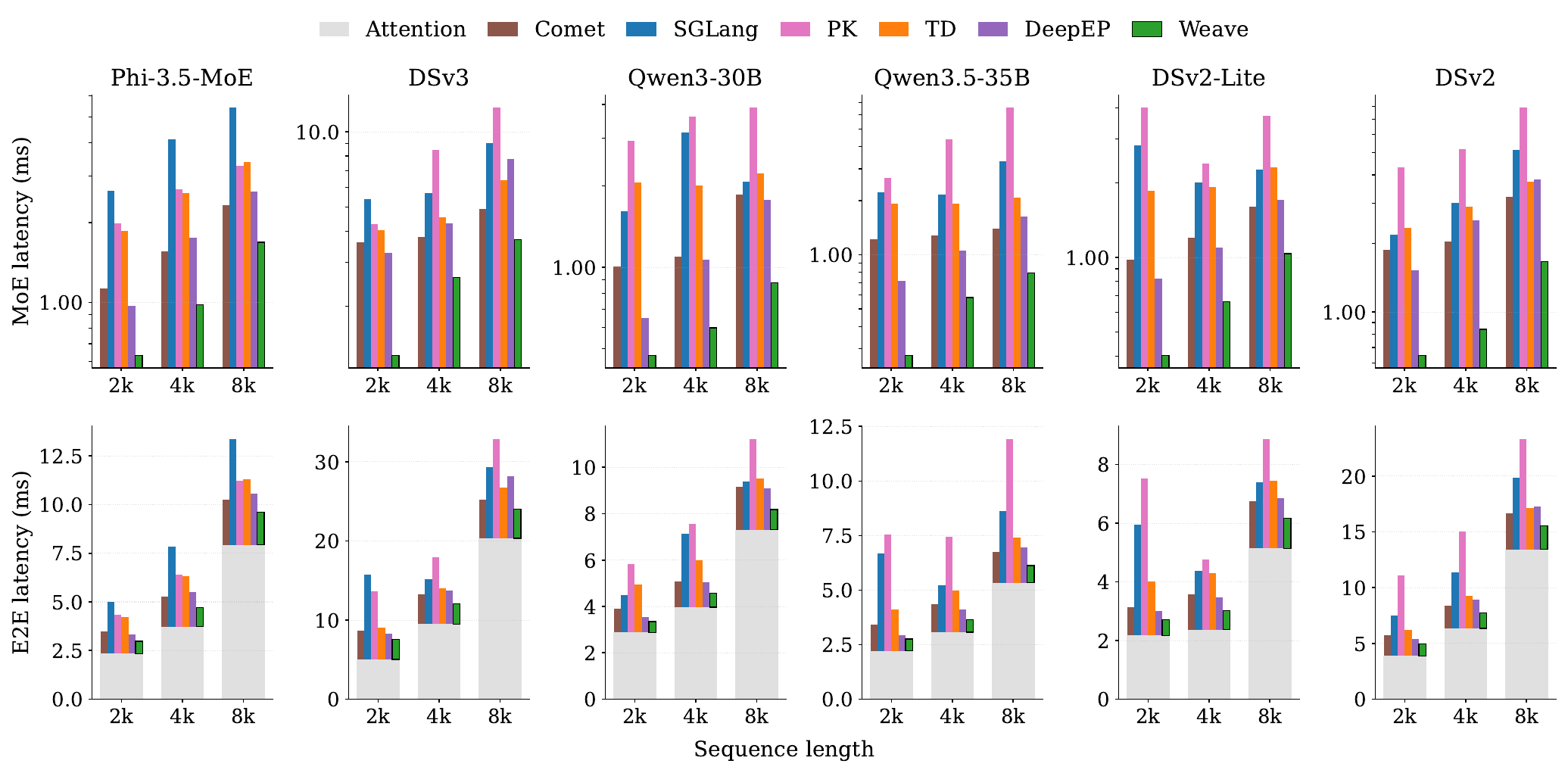}
\caption{Per-layer MoE latency and end-to-end latency on 4$\times$H100 SXM, EP=4, BF16. \textbf{Row 1}: per-layer MoE latency (log-scale y-axis) across six models and three sequence lengths (2k/4k/8k). \textbf{Row 2}: end-to-end latency breakdown for the same six models at the same three sequence lengths; the gray band is attention layer time and the colored bars on top are MoE layer time for each system.}
\label{fig:main-results}
\end{figure*}

We compare Weave against five baselines (SGLang~\cite{sglang}, PK~\cite{parallelkittens}, TD~\cite{tritondist}, DeepEP~\cite{deepep}, Comet~\cite{comet}) across 6 MoE architectures and 3 sequence lengths (2k, 4k, 8k tokens). As shown in Figure~\ref{fig:main-results} (Row 1), Weave achieves the lowest per-layer latency in all 18 configurations.

Weave achieves $1.95\times$--$4.76\times$ geometric mean speedup over all baselines: vs.\ DeepEP $1.95\times$, vs.\ Comet $2.01\times$, vs.\ TD $2.95\times$, vs.\ SGLang $3.63\times$, vs.\ PK $4.76\times$. DeepEP uses DeepGEMM~\cite{deepgemm} as its GEMM backend and chunks tokens into pipeline stages to overlap dispatch/combine with expert GEMM. Comet performs tile-level overlap within two fused kernels (dispatch+gemm0 and gemm1+combine) and can select the communication SM count per iteration based on sequence length. TD fuses the same two operator pairs into two kernels, but within each fused kernel communication and computation run serially across all SMs, with only marginal overlap at operator boundaries via global preemption. PK fuses dispatch with the expert up-projection under a compile-time fixed SM partition. SGLang issues separate NCCL and cuBLAS calls without kernel-level overlap. None of these baselines adapts the SM partition to per-layer routing results.

\subsection{End-to-End Performance}
\label{subsec:e2e}

We evaluate end-to-end (E2E) latency by combining the pre-MoE/attention block time measured via SGLang with the MoE kernel time from Weave and each baseline. We test the same 6 MoE architectures at sequence lengths 2048, 4096, and 8192 (batch size 1, 4$\times$H100 EP=4), yielding 18 configurations per baseline. Figure~\ref{fig:main-results} (Row 2) shows the E2E latency breakdown for all 18 configurations. The gray band represents the attention time and the colored bars show each system's MoE time.

Weave achieves the lowest E2E latency in all 18 configurations, with $1.12\times$--$1.70\times$ geometric mean E2E speedup across baselines: vs.\ DeepEP $1.12\times$, vs.\ Comet $1.13\times$, vs.\ TD $1.28\times$, vs.\ SGLang $1.50\times$, vs.\ PK $1.70\times$. The E2E advantage is more pronounced at shorter sequence lengths: as sequence length grows, attention time scales quadratically while MoE time scales linearly, so the attention portion increasingly dominates E2E latency and dilutes Weave's MoE-layer speedup.

\subsection{Hardware Utilization Profiling}
\label{subsec:nsys}

We profile Weave and other baselines (TD, Comet, DeepEP, PK) with Nsight Systems on DSv2-Lite (seq\,=\,2048, 4$\times$H100 SXM, EP=4). The profiling window covers the complete MoE forward pass from the end of routing to the end of combine. We report three GPU metrics (SM Active, NVLink Utilization, Tensor Core HMMA) and a derived overlap ratio, defined as the fraction of the profiling window during which both NVLink Utilization and Tensor Core HMMA are simultaneously greater than zero.

\begin{figure*}[t]
\centering
\includegraphics[width=0.88\textwidth]{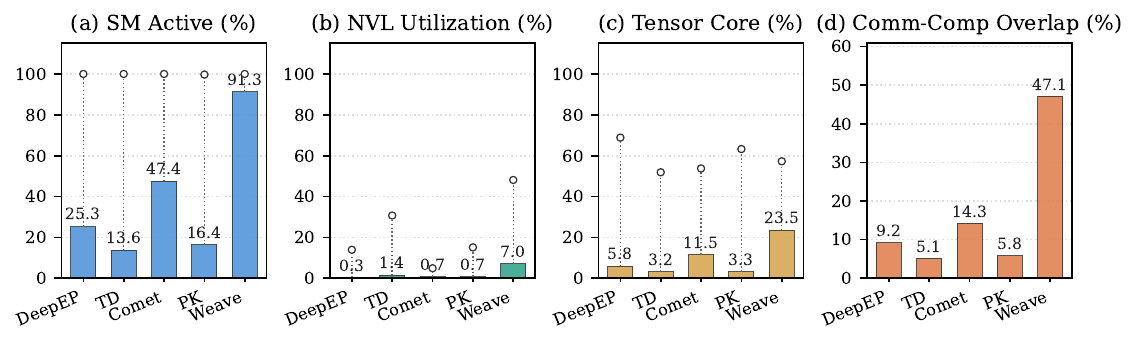}
\caption{Aggregate GPU metrics over the MoE forward window (DSv2-Lite, seq\,=\,2048, 4$\times$H100, EP=4). Bars: mean; hollow circles atop dashed lines: peak. (a)~SM Active (\%). (b)~NVLink Utilization (\%). (c)~Tensor Core HMMA (\%). (d)~Comm-comp overlap ratio (\%).}
\label{fig:nsys-metrics-4panel}
\end{figure*}

Figure~\ref{fig:nsys-metrics-4panel} summarizes the mean (bar) and peak (hollow circle atop dashed line) of the three metrics over each system's MoE forward window, together with the overlap ratio. Weave achieves the highest overlap ratio (47.1\%). The remaining baselines fall well below: Comet (14.3\%), DeepEP (9.2\%), PK (5.8\%), and TD (5.1\%). Comet has the second-highest SM Active mean (47.4\%) because its two-segment fusion keeps SMs busy, but its overlap is moderate. DeepEP's low overlap stems from a mismatched communication SM count: NVLink activity appears as brief bursts that quickly saturate and stop, leaving the vast majority of the timeline with only Tensor Core active.

Weave achieves leading performance across all four metrics. \emph{SM Active}: fusing all five MoE operations into a single persistent megakernel eliminates inter-kernel launch gaps, and bubble stealing ensures communication SMs have GEMM tiles to execute after their primary work finishes, keeping all SMs occupied throughout the layer. \emph{NVLink Utilization} and \emph{Tensor Core HMMA}: the spatial scheduler maintains a well-matched communication/computation SM ratio throughout the kernel's execution, so neither NVLink bandwidth nor compute throughput starves while the other is saturated. \emph{Overlap ratio}: the chunk pipeline extends the communication window from the head and tail of the layer to cover the entire computation phase by interleaving combine with gemm over time, maximizing the duration where NVLink and Tensor Core are simultaneously active.

\subsection{Cost Model Accuracy}
\label{subsec:costmodel-accuracy}

\begin{figure}[!t]
\centering
\includegraphics[width=\linewidth]{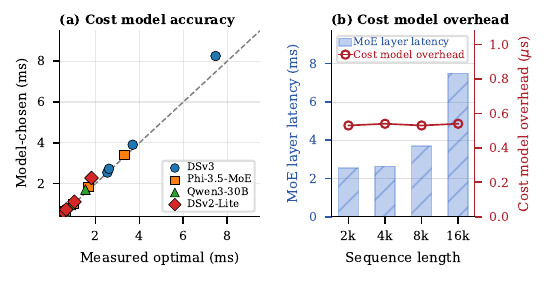}
\caption{Cost model evaluation. (a) Configuration-selection quality: measured-optimal latency vs.\ measured latency at the cost-model-chosen $(c, K)$; dashed line is $y{=}x$. (b) Online overhead on DSv3: MoE layer latency (left blue axis, bars) vs.\ cost model overhead (right red axis, hollow markers).}
\label{fig:cost-model-accuracy}
\end{figure}

We evaluate the cost model along two dimensions: selection accuracy and online cost. As shown in Figure~\ref{fig:cost-model-accuracy}, we sweep 4 models $\times$ 4 sequence lengths on 4$\times$H100 SXM (EP=4), each with 54 $(c, K)$ configurations. \textbf{(a) Selection accuracy.} For each (model, seq\_len) group, we compare the measured latency at the cost-model-selected $(c, K)$ against the measured latency at the exhaustive-search optimum. The average accuracy gap is 8.2\%. (b) Online cost. Because the cost model is fused inside the megakernel, each block computes independently and only reads a small amount of routing results from global memory. On DSv3, the cost model overhead is a nearly constant 0.54\,$\mu$s regardless of sequence length, less than 0.021\% of the MoE layer time, indicating that the runtime search adds negligible overhead.

\subsection{Design Ablation}
\label{subsec:design-ablation}

We isolate the contributions of the spatial and the temporal scheduler by comparing three configurations on Qwen3-30B and Qwen3.5-MoE at sequence lengths 2k, 4k, 8k, and 16k (4$\times$H100, EP=4).
\begin{itemize}
  \item Weave w/o S+T: both schedulers disabled; dispatch / gemm / combine run sequentially on all SMs (no compute-communication overlap).
  \item Weave w/o T: spatial scheduler (S) enabled, temporal scheduler (T) disabled (no chunk pipeline or bubble stealing). Isolates the spatial scheduler's contribution.
  \item Weave: full design with both spatial and temporal schedulers.
\end{itemize}

\begin{figure}[!t]
  \centering
  \includegraphics[width=\linewidth]{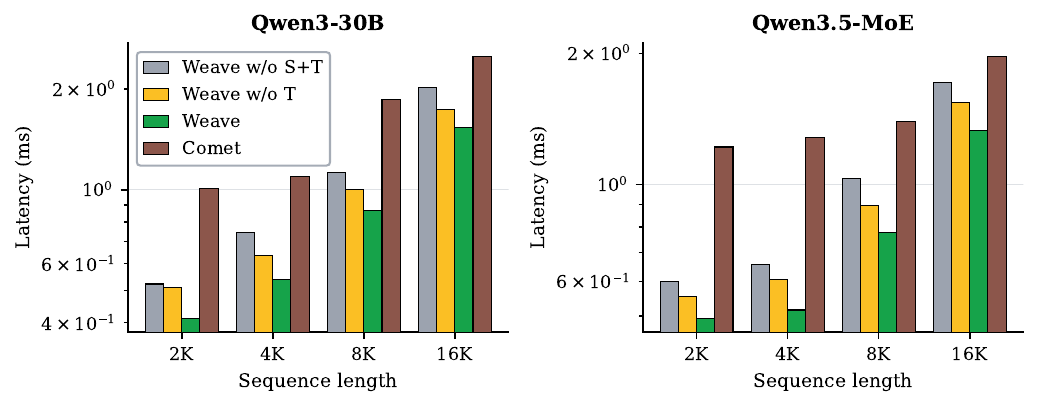}
  \caption{Design ablation on Qwen3-30B and Qwen3.5-MoE (4$\times$H100, EP=4). S\,=\,spatial scheduler, T\,=\,temporal scheduler. Three configurations: Weave w/o S+T (no overlap), Weave w/o T (spatial only), Weave (full).}
  \label{fig:qwen-ablation}
\end{figure}

As shown in Figure~\ref{fig:qwen-ablation}, going from Weave w/o S+T to Weave w/o T quantifies the gain of explicit SM partitioning; going from Weave w/o T to Weave further quantifies the additional gain of the temporal scheduler. Both design components yield significant positive gains at every (model, seqlen) point. On Qwen3-30B at seq=4096, enabling S reduces latency by $14.4\%$, and further enabling T reduces it by $15.7\%$ on top of that; the two stages compose into a $27.8\%$ latency reduction over the no-scheduler configuration. Averaged over the eight (model, seqlen) points, S contributes a $10.1\%$ reduction and T an additional $14.1\%$, for a total $22.8\%$ reduction. This confirms that both schedulers are effective, and their gains stack.

\section{Related Work}
\label{sec:related}

\noindent\textbf{Compute-communication overlap.}
Existing MoE inference systems have extensively studied how to overlap inter-GPU communication with expert computation~\cite{fastermoe,pipemoe,schemoe,lina,lancet,deepseekai2024deepseekv3technicalreport,flowmoe,foldmoe}. Existing systems improve overlap at progressively finer granularity. Host- and stream-level approaches pipeline all-to-all communication with expert computation through operator scheduling and pipeline partitioning~\cite{fastermoe,tutel,pipemoe,schemoe}, but the overlap is still bounded by kernel-level stages.
Finer-grained approaches such as FlashOverlap~\cite{flashoverlap} and TokenWeave~\cite{tokenweave} exploit tile- or token-level opportunities, yet expert computation often remains encapsulated in opaque library kernels, limiting how communication can be interleaved with computation.
Kernel fusion addresses this limitation by moving communication into expert GEMM kernels, allowing dispatch or combine to overlap with GEMM at a finer in-kernel granularity rather than only across separate kernel launches.
ParallelKittens~\cite{parallelkittens} fuses dispatch with the expert up-projection under a compile-time fixed SM partition, while Comet~\cite{comet} and TD~\cite{tritondist} fuse dispatch/up-projection and down-projection/combine as separate stages.
These fused pipelines enable finer-grained overlap by moving communication closer to GEMM execution, but remain limited by fixed SM partitioning, partial fusion, or residual launch boundaries between fused stages.

\noindent\textbf{Persistent megakernel.}
Fusing all operators into a single persistent megakernel has gained increasing attention~\cite{persistent_rnn,rammer,mononn,flashdmoe,mpk,hazy_megakernel,flashformer,event_tensor,uniep}. MoE kernel fusion reduces launch overhead and exposes opportunities for cross-operator scheduling.
Partial-fusion approaches still retain inter-operator boundaries, while whole-layer megakernels such as FlashDMoE~\cite{flashdmoe} integrate dispatch, expert computation, and combine into a single kernel.
Whole-layer fusion removes launch boundaries and creates a larger scheduling domain.
For highly dynamic workloads such as MoE, effectively using this domain requires runtime control over task ordering and resource allocation.
MPK~\cite{mpk} introduces scheduler SMs for device-side task scheduling and also analyzes MoE workloads, but its MoE support is limited to single-GPU execution without cross-GPU dispatch and combine communication. When MPK does consider multi-GPU communication, it targets only static tensor parallelism (TP), thereby missing the compute-communication overlap opportunities unique to expert parallelism where routing-dependent workloads vary per layer.
TD's piecewise megakernel enables limited scheduling within fused regions, but still only partially fuses MoE operators and does not jointly adapt SM partitioning and intra-layer execution order after routing.
Weave addresses this gap by using a persistent megakernel as a routing-aware scheduling domain that adapts both decisions at runtime.

\section{Conclusion}
\label{sec:conclusion}

We present Weave, to our knowledge the first MoE overlap system whose SM split is decided per layer and per GPU by routing results at runtime. Existing systems either fix the SM partition at compile time, adjust it only at coarse granularity, or avoid explicit SM partitioning and run communication and computation serially with minimal overlap; Weave instead exploits the fact that each layer's communication and computation volumes become known after routing, running a lightweight cost model inside the persistent megakernel to jointly select the communication SM count and chunk pipeline configuration online. On 4$\times$H100 SXM GPUs across six mainstream MoE models, Weave achieves a $2.89\times$ geometric-mean MoE-layer speedup and a $1.33\times$ geometric-mean end-to-end speedup over five state-of-the-art baselines.

\textbf{Limitations and future work.}
Due to hardware constraints, Weave is currently validated on a single 4$\times$H100 NVLink node with EP=4. Extending to larger GPU counts and multi-node configurations is left for future work.

\bibliographystyle{ACM-Reference-Format}
\bibliography{refs}

\end{document}